\documentclass{PoS}

\title{The gradient flow in a twisted box}
\usepackage{multirow}
\usepackage{graphicx}
\usepackage{subfig}

\ShortTitle{The gradient flow in a twisted box}

\author{\speaker{Alberto Ramos}\\
        NIC, DESY. Platanenallee 6, 15738 Zeuthen, Germany\\
        E-mail: \email{alberto.ramos@desy.de}}

\abstract{We study the perturbative behavior of the gradient flow in a
twisted box. We apply this information to define a running
coupling using the energy density of the flow field. 
We study the step-scaling function and the size of cutoff effects in
SU(2) pure gauge theory. We conclude that the twisted gradient flow
running coupling scheme is a valid strategy for step-scaling purposes
due to the relatively mild cutoff effects and high precision.
\vspace{2cm}
\begin{flushright}
DESY 13-108
\end{flushright}
}

\FullConference{31st International Symposium on Lattice Field Theory - LATTICE 2013\\
		July 29 - August 3, 2013\\
		Mainz, Germany}

\begin{document}

\section{Introduction}

The Yang-Mills gradient flow~\cite{Luscher:2010iy,Luscher:2011bx} is a new
powerful tool to investigate non-perturbative aspects of non abelian
quantum field theories. In the context of $SU(N)$ Yang-Mills theories
one introduces an extra coordinate $t$ (called
flow time, not the same as Euclidean time, denoted $x_4$), and defines
a flow gauge field $B_\mu(x,t)$ according to the equation
\begin{eqnarray}
  \label{eq:flow}
  \frac{ {\rm d}B_\mu(x,t)}{{\rm d}t} &=& D_\nu G_{\nu\mu}(x,t)\,, \\
  G_{\mu\nu} &=& \partial_\mu B_\nu - \partial_\nu B_\mu + 
  [B_\mu,B_\nu] \,, 
\end{eqnarray}
with initial condition
\begin{equation}
  B_\mu(x,0) = A_\mu(x) \,.
\end{equation}

Since $D_\nu G_{\nu\mu}(x,t) \sim -\delta S_{YM}/\delta B_{\mu}$ the
flow drives the gauge field towards a classical solution (i.e. the
flow smooths the field). Due to this smoothing property ultra-violet
fluctuations are suppressed and composite operators do not need any
renormalization at positive flow time. In particular the energy
density
\begin{equation}
  \langle E(t) \rangle = \frac{1}{4}\langle G_{\mu\nu}G_{\mu\nu}\rangle
\end{equation}
has the following perturbative expansion in $\mathbb
R^4$\cite{Luscher:2010iy} 
\begin{equation}
    \langle E(t)\rangle =
    \frac{3g^2_{\overline{MS}}}{16\pi^2t^2}(1+c_1g^2_{\overline{MS}}+\mathcal O(g^4_{\overline{MS}}))
\end{equation}
and therefore can be used for a non-perturbative definition of the 
coupling at a scale $\mu=1/\sqrt{8t}$
\begin{equation}
  \alpha(\mu) = \frac{4\pi}{3}t^2\langle E(t)\rangle = 
  \alpha_{\overline{MS}}(\mu) + \dots\,.
\end{equation}
Moreover, in a finite volume one can identify the renormalization
scale $\mu$ with the size of the system and use the energy density to
define a running coupling. 

In these proceedings we would like to explore an alternative to the
previously studied running coupling schemes based on the
gradient flow that use either periodic~\cite{Fodor:2012td} or Schr\"odinger
Functional~\cite{Fritzsch:2013je} boundary conditions. It 
is based on twisted boundary conditions for the gauge fields and
has several practical advantages. The twisted boundary conditions, if
chosen appropriately,
guarantee that the only zero modes of the action are the gauge modes,
and therefore this coupling definition is analytic in
$g^2_{\overline{\rm MS}}$ and has an universal 2-loop beta
function. Moreover in
this setup the fields still live on a torus and therefore 
there are no boundary counterterms: $\mathcal O(a)$ improvement is
guaranteed without any tuning. The weak
point of this running coupling scheme shows up when one considers
Yang-Mills fields coupled to matter. Fermions in the fundamental
representation are naively incompatible with the twisted boundary
conditions. On the other hand
fermions in multi-index representations (like adjoint fermions) do not
suffer from this obstruction.

\section{Twisted boundary conditions}

We have no space here to review the basics of twisted boundary
conditions. Instead we will say a few basic things to clarify the
notation and refer the reader to the
literature. In~\cite{Perez:2013dra} the reader can find a nice summary
on twisted boundary conditions with a similar
notation and in a similar setup than the ones used here. This setup is
basically the same as the one used in the Twisted Polyakov Loop (TPL)
scheme~\cite{deDivitiis:1994yp}. Finally the review~\cite{ga:torus}
contains more in-depth information and proofs about some of the
results used here.  

We are going to work in a four dimensional torus $\mathcal T^4$ of
sides $L\times L\times L\times L$. The basic idea is that on a torus
physical quantities have to be 
periodic, but this does not imply that the gauge potential $A_\mu(x)$
has to be periodic. It is enough if it is periodic modulo a gauge
transformation
\begin{equation}
  A_\mu(x+L\hat \nu) = \Omega_\nu(x)A_\mu(x)\Omega^+_\nu(x) + 
  \Omega_\nu(x)\partial_\mu \Omega_\nu^+(x).
\end{equation}

The matrices $\Omega_{\mu}(x)$ are called twist matrices, and they
have to obey the consistency relation
\begin{equation}
  \Omega_\mu(x+L\hat\nu)\Omega_\nu(x) = z_{\mu\nu}
  \Omega_\nu(x+L\hat\mu)\Omega_\mu(x) 
\end{equation}
where $z_{\mu\nu}$ are elements of the center of $SU(N)$. The
particular choice of twist matrices is irrelevant, since they change
under gauge transformations, but it is easy to check that $z_{\mu\nu}$
is gauge invariant, and therefore it encodes the physical part of the
twisted boundary conditions. We are going to use a particular setup of
this general scheme: we twist the plane $x_1,x_2$, while
the gauge potential will be periodic in the directions $x_3$ and
$x_4$. Moreover we are going to choose 
\begin{equation}
  z_{12} = z = e^{2\pi\imath/N},
\end{equation}
and the twist matrices to be constant $\Omega_{1,2}(x) = \Omega_{1,2}$
obeying the relation
\begin{equation}
  \Omega_1\Omega_2 = e^{2\pi\imath /N}
  \Omega_2\Omega_1.
\end{equation}

We will define the usual space momentum 
\begin{equation}
  p_\mu = \frac{2\pi n_\mu}{L}\,, \quad 
  \mu = 1,2,3,4;\; n_\mu = 0,\dots
\end{equation}
and what is usually called the color momentum 
\begin{equation}
  \tilde p_i = \frac{2\pi \tilde n_i}{NL}\,, \quad 
  i = 1,2;\; \tilde n_i = 0,\dots,N-1.
\end{equation}
The total momentum is the sum of both
\begin{eqnarray}
  P_{i} &=& p_i + \tilde p_i \quad (i=1,2)\,, \\
  P_{3,4} &=& p_{3,4} \,.
\end{eqnarray}
It can be proved~\cite{ga:torus} that any gauge connection compatible
with these particular boundary conditions can be written as
\begin{equation}
  A_\mu^a(x)T^a = \frac{1}{L^4} \sum_{p,\tilde p\ne 0}\tilde
  A_\mu(P)e^{\imath  
    Px}\hat\Gamma(P),
\end{equation}
where $\tilde A_\mu(P)$ are complex coefficients (\emph{not}
matrices) and $\hat\Gamma(P)$ are matrices given by
\begin{equation}
    \hat\Gamma(P) = \Omega_1^{-k\tilde
    n_2}\Omega_2^{k\tilde n_1}.
\end{equation}

Note that the only constant gauge connection compatible with our choice
of boundary conditions is $A_\mu = 0$, and this is the
only configuration (up to gauge transformations) with zero action. 

\section{$\langle E(t) \rangle$ to leading order in a twisted box and
  running coupling definition} 

In perturbation theory one scales the gauge potential with the bare
coupling $A_\mu \rightarrow g_0 A_\mu$. The flow
field of equation~(\ref{eq:flow}) becomes a function of $g_0$ with 
an asymptotic expansion
\begin{equation}
  B_\mu(x,t) = \sum_{n=1}^{\infty} B_{\mu,n}(x,t)g_0^n.
\end{equation}
After gauge fixing and inserting this expansion in~(\ref{eq:flow}), we
find that to leading order the gradient flow equation reads
\begin{equation}
   \frac{ {\rm d}B_{\mu,1}(x,t)}{{\rm d}t} = 
   \partial_\nu^2 B_{\mu,1}(x,t)\,, \quad
   B_{\mu,1}(x,0) = A_\mu(x).
\end{equation}

The solution to this linear equation compatible with our twisted
boundary conditions can be written as
\begin{equation}
  \label{eq:sol}
  B_{\mu,1}(x,t) = \frac{1}{L^4}
  \sum_{p,\tilde p\ne 0} e^{-P^2t} \tilde A_\mu(P) e^{\imath Px}
  \hat\Gamma(P).
\end{equation}

We will expand our observable of interest $\langle E(t) \rangle$ in
powers of $g_0$
\begin{equation}
  \langle E(t)\rangle = \frac{1}{4}\langle
  G_{\mu\nu}(x, t)G_{\mu\nu}(x,t)\rangle = 
  \mathcal E(t) + \mathcal O(g_0^4)\,,
\end{equation}
where the leading order contribution is given by
\begin{equation}
  \mathcal E(t) = \frac{g_0^2}{2}\langle 
  \partial_\mu B_{\nu,1}\partial_\mu B_{\nu,1} - \partial_\mu
  B_{\nu,1}\partial_\nu B_{\mu,1} 
  \rangle.
\end{equation}
A short computation gives as result
\begin{equation}
  \mathcal E(t) = 
  \frac{3g_0^2}{2L^4}\sum_{p,\tilde p\ne 0} e^{-P^2t}\,.
\end{equation}

To define a running coupling we simply identify the renormalization scale
$\mu = 1/\sqrt{8t}$ with the linear size of the finite volume box
\begin{equation}
  \sqrt{8t} = cL\,.
\end{equation}
The parameter $c$ identify the scheme. The definition of the
twisted gradient flow running coupling reads
\begin{equation}
    g_{TGF}^2(L) = \mathcal N^{-1}_T(c)t^2\langle E(t) \rangle \Big|_{t=c^2L^2/8}
  = g_{\overline{\rm MS}}^2 + \mathcal O(g_{\overline{\rm MS}}^4)
\end{equation}
with
\begin{equation}
      \mathcal N_T(c) =
  \frac{3g_0^2c^4}{128}\sum_{p,\tilde p\ne 0} e^{-\frac{c^2L^2}{4}P^2}
  = \frac{3g_0^2c^4}{128}\sum_{n_\mu=-\infty}^{\infty}
  {\sum_{\tilde n_i=0}^{N-1}}' 
  e^{-{\pi^2c^2}(n^2 + \tilde n^2/N^2 + 2\tilde n_i n_i/N)}\,.
\end{equation}
The prime over the sum recalls that the term $\tilde n_1 =
\tilde n_2 = 0$ has to be dropped. 

\section{$SU(2)$ running coupling}

As a test we have performed a non-perturbative running of the
$SU(2)$ coupling. Here we will only sketch the simulations and
results. A detailed description of the simulations, algorithms and
parameters will be presented later~\cite{toappear}. We have performed
simulations with the Wilson action for $\beta \in [2.75,12.0]$ on
lattices of sizes $L/a = 10, 12, 15, 18$, and to 
perform the step-scaling, also on lattices of sizes $L/a = 20, 24, 30,
36$. We collect 2048 measurements of twisted gradient flow
coupling for $c=0.3$. The measurements are well spaced in simulation
time and autocorrelations are negligible. With this statistics we
achieve a precision between a $0.15-0.3\%$, independently of the value
of $L/a$ (see table~\ref{tab:gsq} for some representative values of
the coupling). 

\begin{table}
  \centering
  
  \begin{tabular}{l|llll}
    $\beta$ & \multicolumn{3}{c}{$g^2_{\rm TGF}(L)$} \\
    \hline
    & $L/a=20$ & $L/a=24$& $L/a=30$& $L/a=36$ \\
    \hline
    12.0 & 0.40162(60)&0.40514(62)&0.40772(64)& 0.41078(64)  \\
    10.0 & 0.50340(78)&0.50786(79)&0.51280(81)& 0.51809(83) \\
    8.0 &  0.6742(11) &0.6809(11) &0.6900(11) & 0.6987(11)  \\
    7.0 &  0.8125(13) &0.8240(13) &0.8369(13) & 0.8497(13)  \\
    6.0 &  1.0222(17) &1.0379(17) &1.0608(17) & 1.0819(18) \\
    5.0 &  1.3766(23) &1.4091(23) &1.4562(25) & 1.4968(25) \\
    4.0 &  2.1439(38) &2.2260(41) &2.3453(42) & 2.4465(44) \\
    3.75 & 2.5047(47) &2.6107(47) &2.7636(51) & 2.9277(54) \\
    3.5  & 3.0037(57) &3.1720(60) &3.4170(65) & 3.6494(69) \\
    3.25 & 3.7581(72) &4.0223(76) &4.4397(87) & 4.8568(99) \\
    3.0 &  5.088(11)  &5.630(12)  &6.573(15)  & 7.587(20)  \\
    2.9 &  5.989(13)  &6.835(17)  &8.518(26)  & -   \\
    2.8 &  7.439(20)  &9.116(27)  &-  & -  \\
    2.75&  8.699(25)  &-  & - & -  \\
    \hline
  \end{tabular}

%%% Local Variables: 
%%% mode: latex
%%% TeX-master: "lat13"
%%% End: 

  \caption{Values of the twisted gradient flow coupling for different
    values of $\beta$ in lattices of sizes $L/a=20, 24, 30, 36$. These
    values were determined with 2048 measurements well spaced in Monte
  Carlo time so that autocorrelations were negligible.}
  \label{tab:gsq}
\end{table}

\begin{figure}
  \centering
  \subfloat[][]{
  \includegraphics[width=0.45\textwidth]{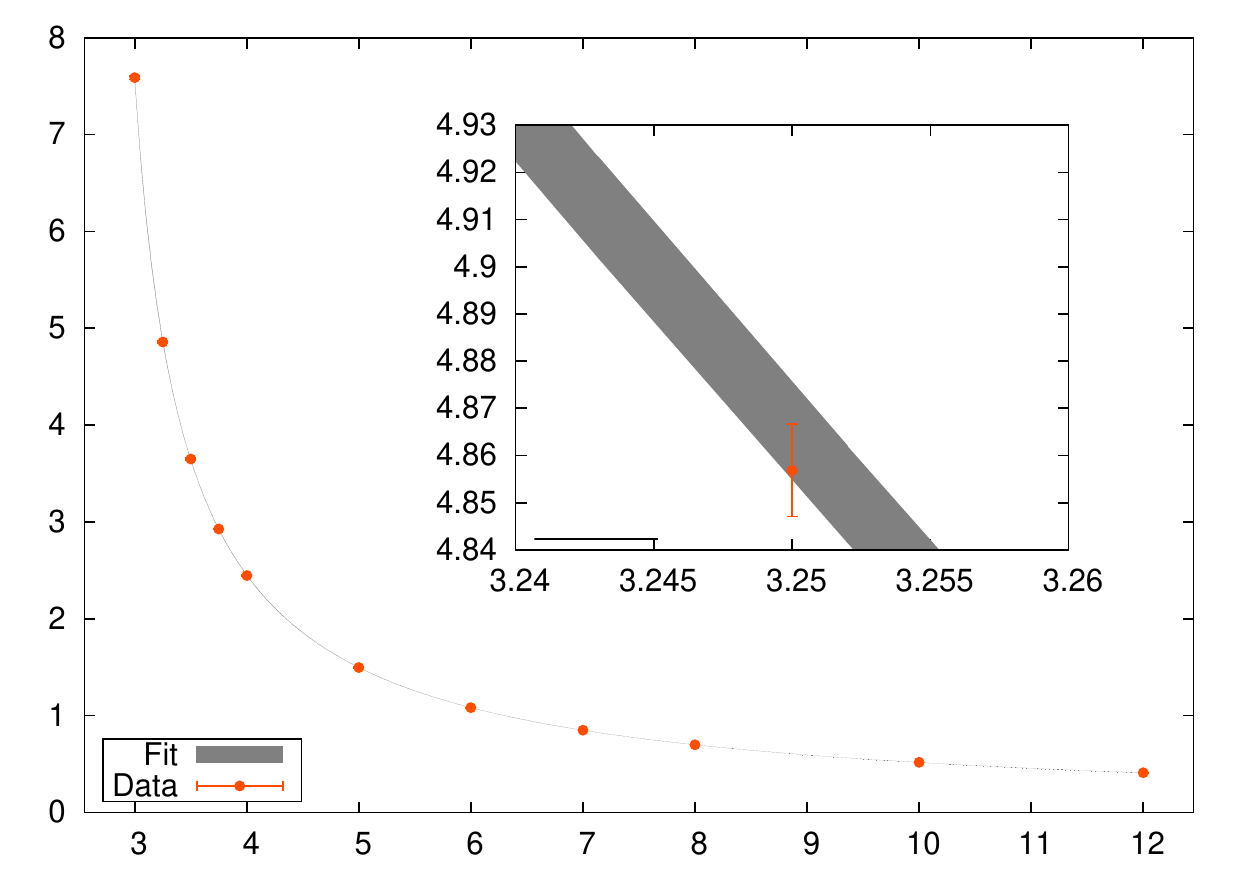}
  \label{fig:fit}
}
  \subfloat[][]{
  \includegraphics[width=0.45\textwidth]{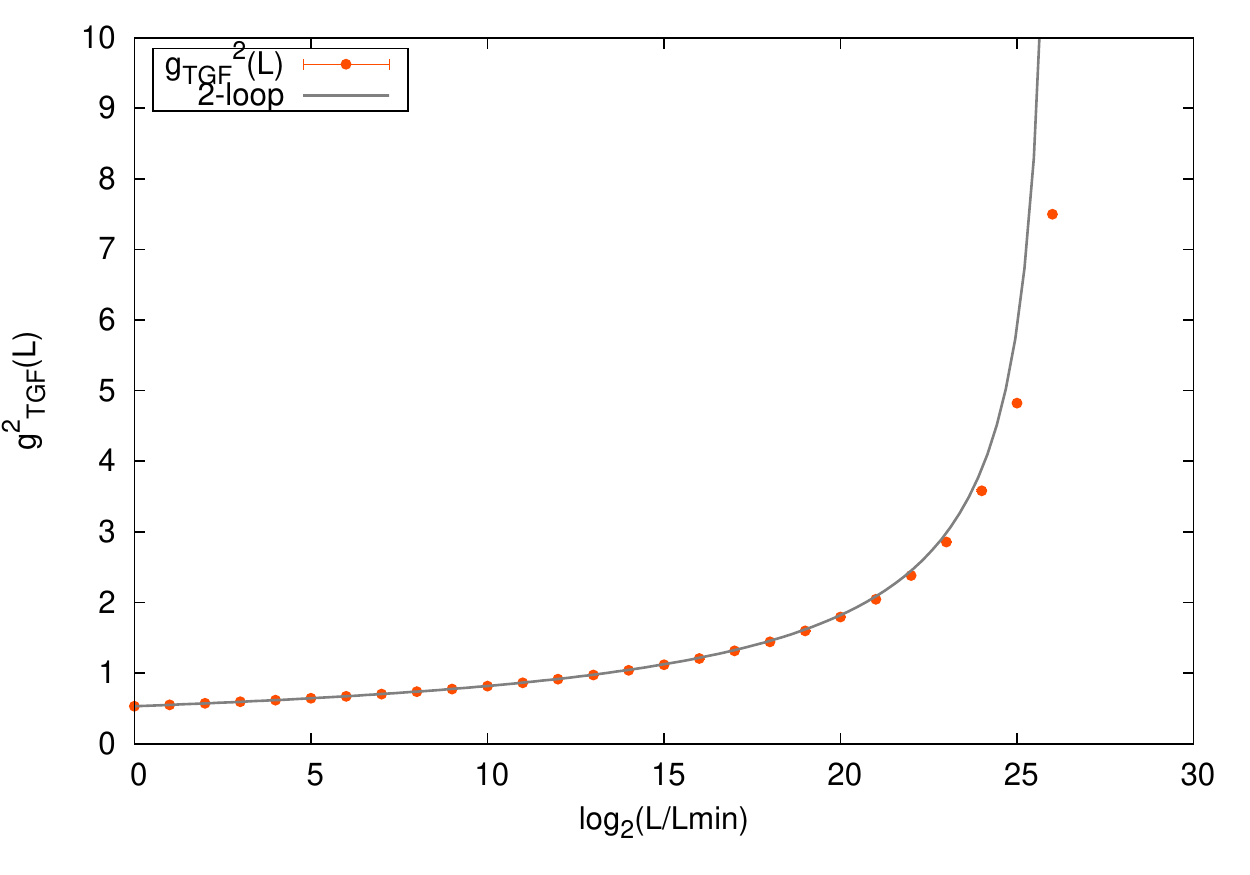}
  \label{fig:gvsL}
}
  \caption{Left: The twisted gradient flow coupling for $L/a=36$ as a
    function of $\beta$. The points are fitted to a Pad\'e functional
    form (grey band in the figure) with 4 parameters and the
    $\chi^2/{\rm ndof} = 5.9/7$. Right: Non-perturbative running of
    the twisted gradient flow coupling.}
\end{figure}

For each value of $L/a$ we fit the data using a Pad\'e ansatz of the
form
\begin{equation}
  g^2_{TGF}(a/L,\beta) = \frac{4}{\beta} \frac{\sum_{n=0}^{N-1} a_n(a/L)\beta^n
  + \beta^N}{\sum_{n=0}^{N-1} b_n(a/L)\beta^n+\beta^N}.
\end{equation}
We obtain fits with good quality ($\chi^2/{\rm ndof}\sim 1$)
with 4 parameters in all our cases. An example of such a fit, for the
case of the $L/a=36$ lattice can be seen in the figure~\ref{fig:fit}.
\begin{figure}
  \centering
  \subfloat[][]{
    \includegraphics[width=0.3\textwidth]{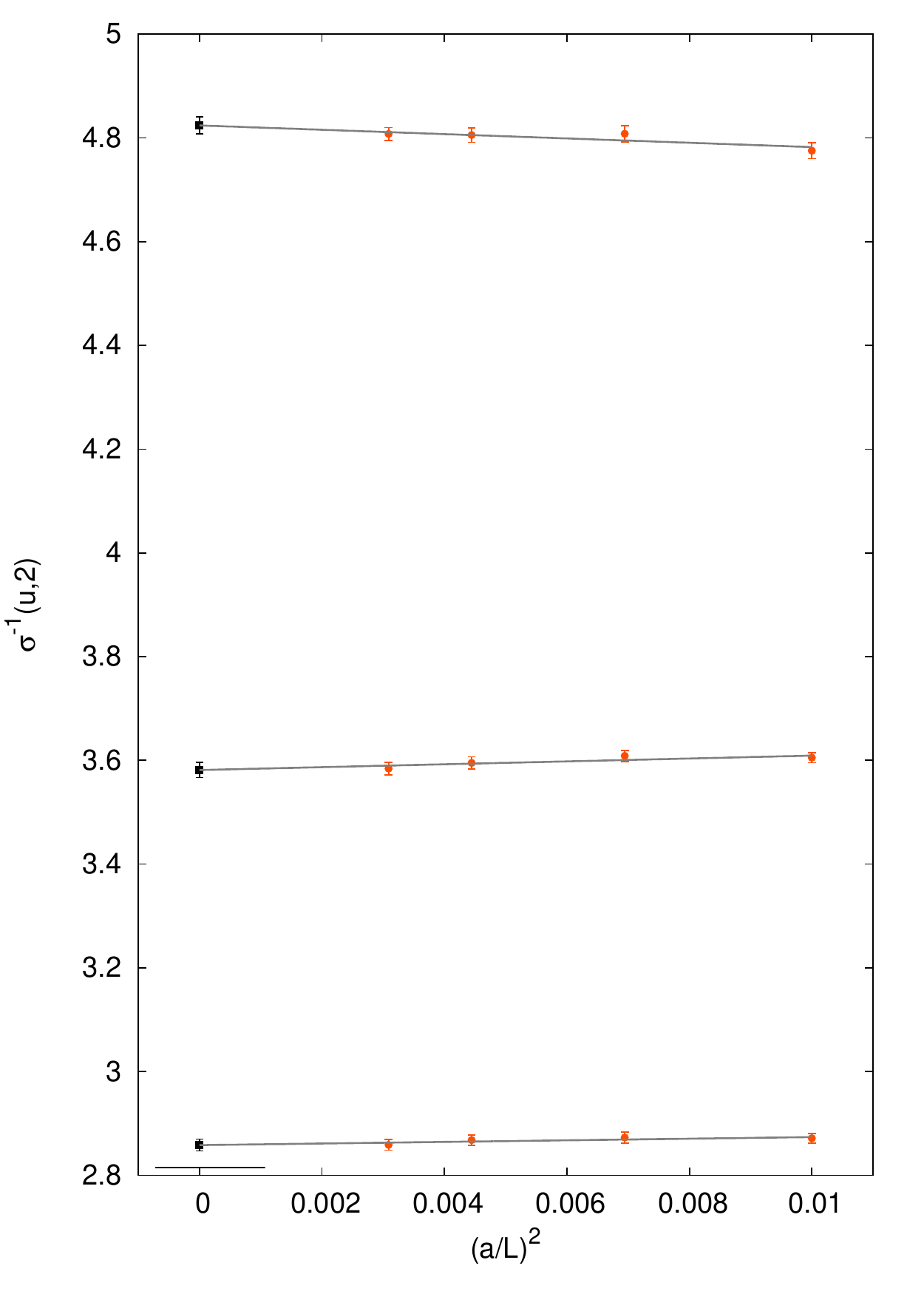}
  }
  \subfloat[][]{
    \includegraphics[width=0.3\textwidth]{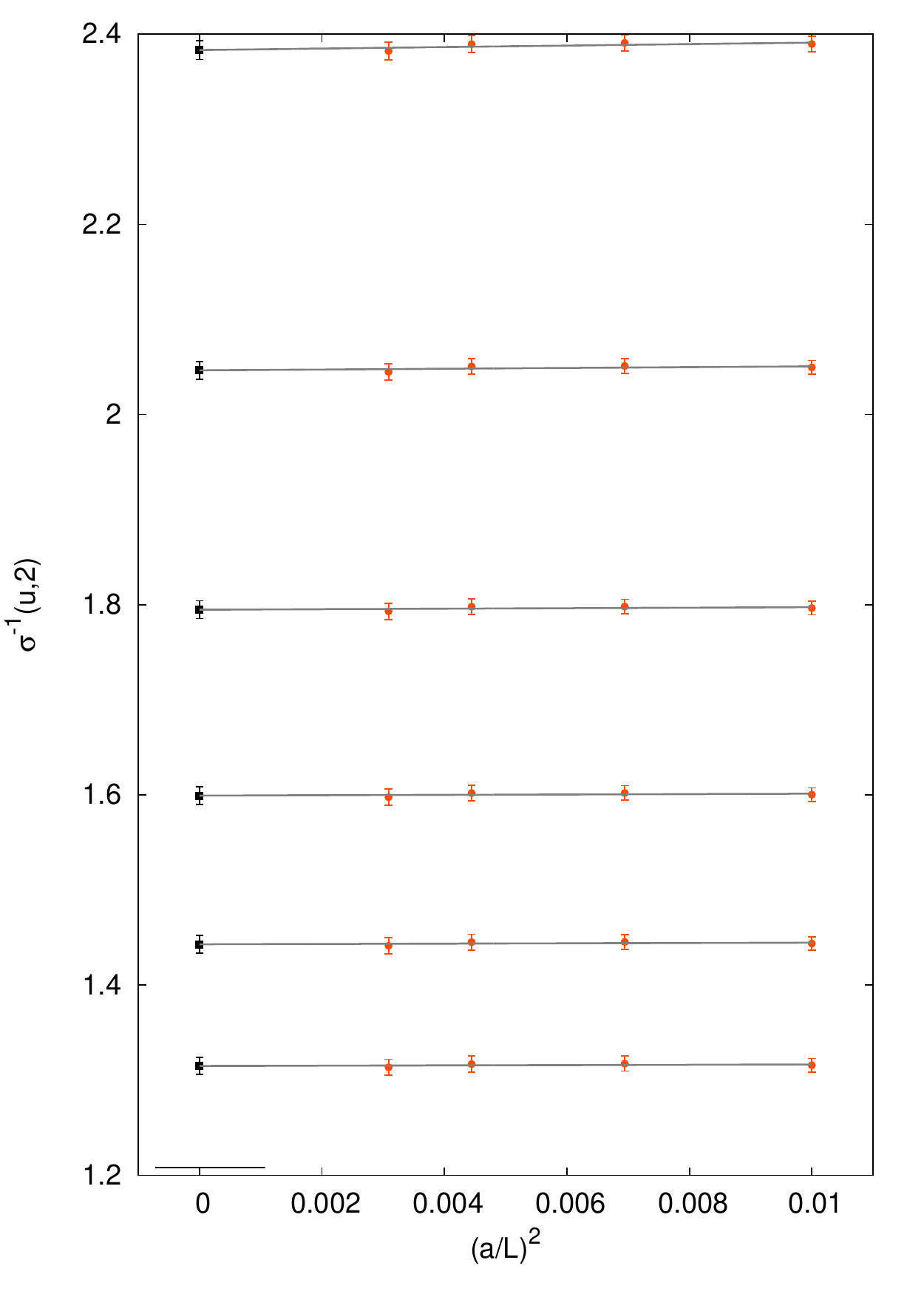}
  }
  \subfloat[][]{
    \includegraphics[width=0.3\textwidth]{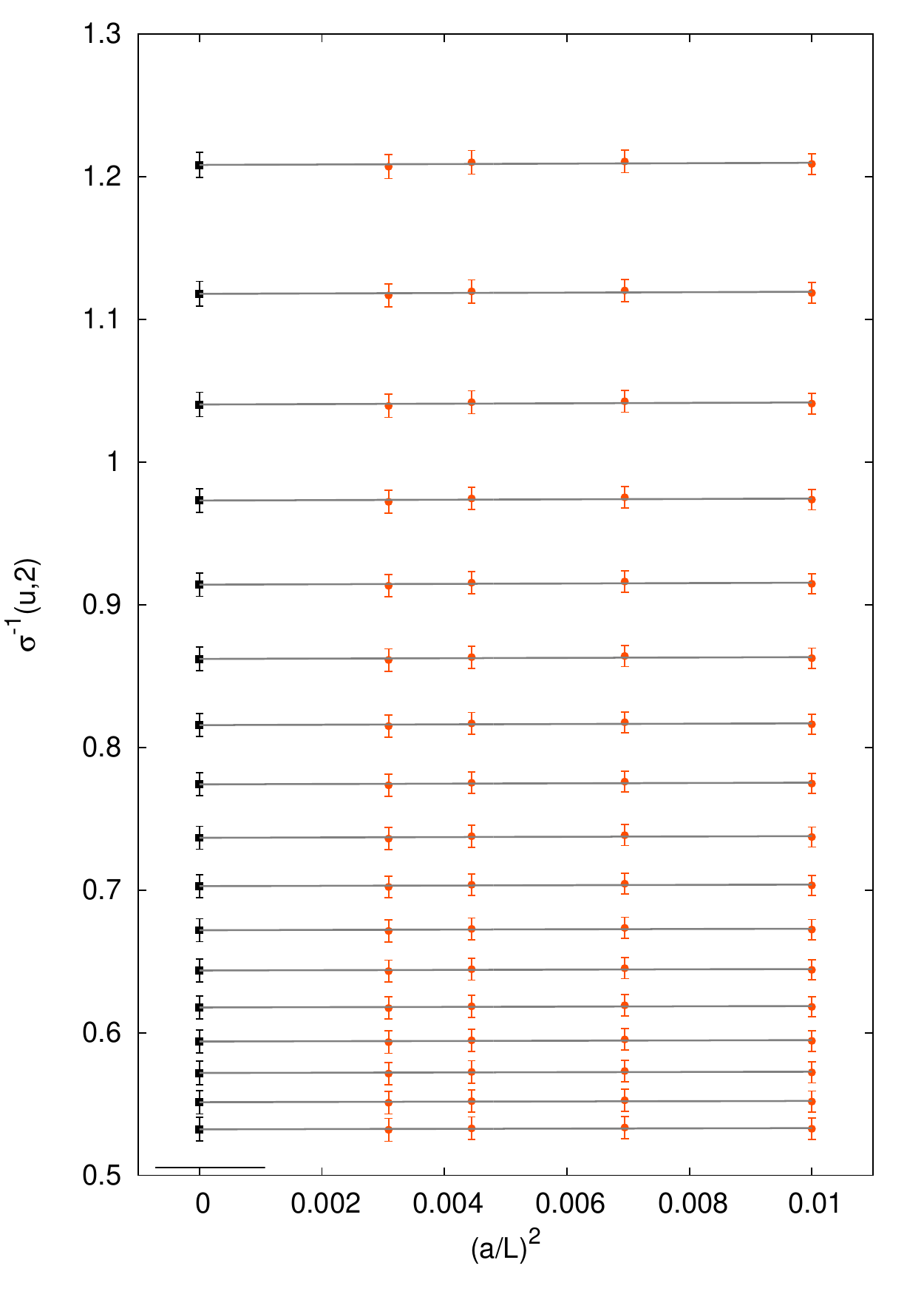}
  }
  \caption{Continuum extrapolation of the step-scaling function. We
    start the recursion in a volume $L_{\rm max}$ where $g^2_{\rm
      TGF}(L_{\rm max})=7.5$.} 
  \label{fig:ss}
\end{figure}

We apply the usual step-scaling technique, starting the recursion in a
volume $L_{\rm max}$ where $g^2_{\rm TGF}(L_{\rm max})=7.5$. The
continuum extrapolations of the step-scaling function are very flat,
as can be seen in figure~\ref{fig:ss}. 
The non-perturbative
running of the coupling is recursively carried over a factor $2^{25}$
change in the scale down to a volume $L_{\rm min}$ where 
$g^2_{\rm TGF}(L_{\rm min})=0.5323(83)$. Figure~\ref{fig:gvsL} shows
the scale dependence of the coupling and a comparison with
the universal 2-loop $\beta-$function.

\section{Conclusions and comments}

In this proceedings we have studied perturbatively the gradient flow
in a four dimensional torus with twisted boundary conditions. The
energy density of the flow field can be used to define a running
coupling. We have presented some preliminary results on a $SU(2)$
running coupling where we have shown that the twisted gradient flow
coupling is a convenient choice. The observable is precise and 2048
independent measurements are enough to reach a per mile
precision. Cutoff effects of the step-scaling functions are mild.  

Some comments about the inclusion of matter
fields are in order. In principle our coupling definition is perfectly
valid with 
any kind of matter fields, but fermions in the fundamental
representation are incompatible with the twisted boundary
conditions. A fermion in the fundamental representation transforms as
\begin{equation}
  \psi(x+L\hat\mu) = \Omega_\mu\psi(x)\,,
\end{equation}
but this transformation is not consistent due to
\begin{eqnarray}
  \psi(x+L\hat 1+L\hat 2) &=& \Omega_1\Omega_2\psi(x)\\
  \psi(x+L\hat 2+L\hat 1) &=& \Omega_2\Omega_1\psi(x)
  = e^{2\pi\imath/N} \Omega_1\Omega_2\psi(x)\,.
\end{eqnarray}
This obstruction can be overcome if the number of fermions is an
integer multiple of the degree of the gauge group
$N$~\cite{Parisi:1984cy}.  

On the other hand fermions in multi-index representations (like
adjoint fermions) do not suffer from any kind of restriction.

\section*{Acknowledgments}

I want to thank M. Garc\'ia P\'erez and A. Gonz\'alez-Arroyo
for sharing some of their results before publication and A. Sastre for
the help in testing some parts of the codes. The continuous help of
R. Sommer in each of the steps of this work was invaluable.

I also want to thank P. Fritzsch, P. Korcyl, H. Simma, S. Sint and
U. Wolff for the many useful discussions on this work.

\bibliography{/home/alberto/docs/bib/math,/home/alberto/docs/bib/campos,/home/alberto/docs/bib/fisica,/home/alberto/docs/bib/computing}

\end{document}